\documentclass[aas_macros]{aa}

\usepackage{graphicx}
\usepackage{natbib}
\def\chandra{{\it Chandra~}}
\def\integral{{\it INTEGRAL~}}
\def\gm{{$\gamma$-}}

\title{INTEGRAL detection of the pulsar wind nebula in \hbox{PSR~J1846$-$0258}}
\author{V.A.~McBride\thanks{vanessa@soton.ac.uk}\inst{1} \and A.J.~Dean\inst{1} \and A.~Bazzano\inst{2} \and A.J.~Bird\inst{1} \and A.B.~Hill\inst{1} \and A.~De~Rosa\inst{2} \and R.~Landi\inst{3} \and V.~Sguera\inst{3} \and A.~Malizia\inst{3}}

\authorrunning{V.A. McBride et al.}
\titlerunning{INTEGRAL detection of PSR J1846$-$0258}

\institute{School of Physics \& Astronomy, University of Southampton, Highfield, SO17 1BJ, United Kingdom \and INAF/IASF, via del Fosso del Cavaliere 100, Roma, 00113, Italy
\and INAF/IASF, via P. Gobetti 101, Bologna, 40129, Italy}

\begin{document}

\date{Received 7 August 2007 / Accepted 26 October 2007}

\abstract{}{We communicate the detection of soft (20--200\,keV) $\gamma$-rays from the pulsar and pulsar wind nebula of \hbox{PSR~J1846$-$0258} and aim to identify the component of the system which is responsible for the $\gamma$-ray emission.}{We combine spectral information from the \integral $\gamma$-ray mission with archival data from the \chandra X-ray Observatory to pinpoint the source of soft $\gamma$-ray emission.}{Our analysis shows that the soft $\gamma$-rays detected from \hbox{PSR~J1846$-$0258} include emission from both the pulsar and the pulsar wind nebula, but the measured spectral shape is dominated by the pulsar wind nebula.  We discuss \hbox{PSR~J1846$-$0258} in the context of rotation powered pulsars with high magnetic field strengths and review the anomalously high fraction of spin-down luminosity converted into X- and $\gamma$-ray emission in light of a possible overestimate of the distance to this pulsar.}{}


\maketitle

\section{Introduction}

The pulsar \hbox{PSR~J1846$-$0258} (also known as \hbox{AX~J1846.4$-$0258}) was discovered by \citet{GotthelfVasishtBoylan-Kolchin2000} in the X-ray band and lies near the centre of the supernova remnant Kes 75 (SNR G29.7-0.3).  Recent X-ray imaging \citep{HelfandCollinsGotthelf2003} shows that the pulsar is embedded in a pulsar wind nebula (PWN) which shows distinct physical structure.  

No radio emission has been observed from \hbox{PSR~J1846$-$0258}, but X-ray timing reveals the pulse period to be $P=324$\,ms and the characteristic age is $P/(2\dot{P})=728$\,years -- the smallest characteristic age of any rotation-powered pulsar.  The inferred surface dipole magnetic field strength is $4.8\times10^{13}$\,G -- almost an order of magnitude greater than the magnetic fields for typical pulsars, placing it closer to the field strengths of magnetars. 

The distance to the system is $\sim19$\,kpc as estimated from the neutral hydrogen density along the line of sight \citep{BeckerHelfand1984}, which implies that the size of the supernova remnant (SNR) is extremely large for such a young pulsar.  It also implies that the efficiency of the conversion of the pulsar spin-down luminosity ($8.4\times10^{36}$\,erg\,s$^{-1}$, using $P$ and $\dot P$ from \citealt{GotthelfVasishtBoylan-Kolchin2000}) to combined pulsar and PWN X-ray luminosity in the 0.5--10\,keV energy range, is 20\% (using the flux from this work) -- the largest of any rotation-powered pulsar. 

 This source is one of a growing class of rotation-powered pulsars with B-field strengths approaching those of magnetars.  In this paper we present the \integral observations and results for \hbox{PSR~J1846$-$0258}, link these results to previously published \chandra results and follow up with a discussion of the properties of this source in comparison to other members of this class and examine the possibility of an overestimate in the distance to this pulsar.

\begin{figure*}
\sidecaption
\includegraphics[width=12cm]{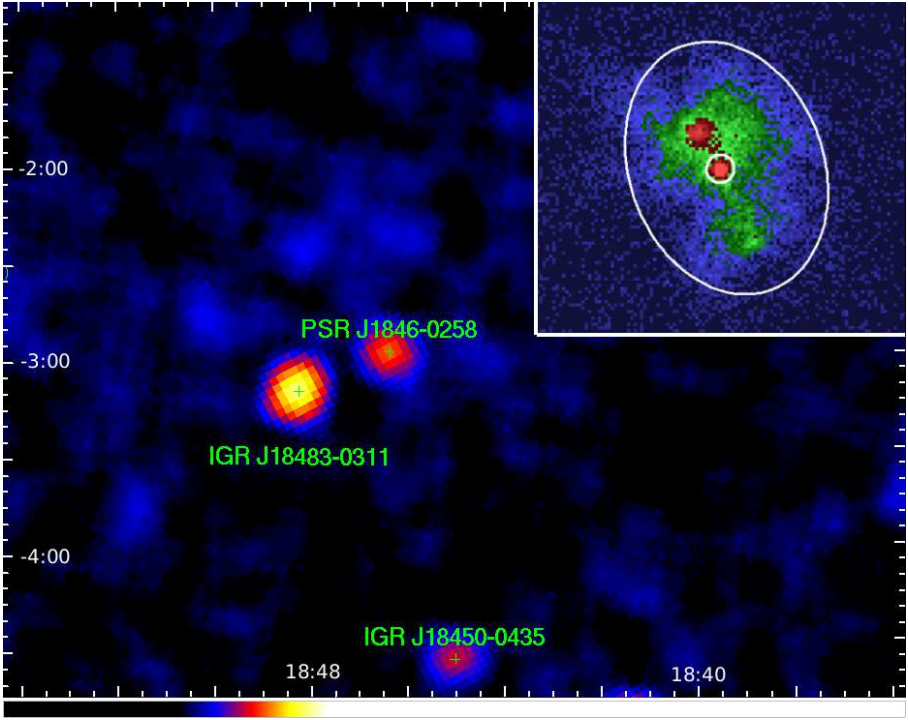}
\caption{IBIS/ISGRI significance mosaic of the 20--100\,keV energy band, showing \hbox{PSR~J1846$-$0258} in the centre of the field.  False colour representation of the significance is displayed on a logarithmic scale.  The top right inset is a $50\arcsec\times50\arcsec$ image from our reduction of \chandra data in the 0.3--10\,keV energy band.  One can clearly distinguish the bright pulsar and the surrounding synchrotron nebulosity.  The white ellipses on the \chandra image indicate the extraction regions for the PWN and pulsar spectra as explained in Sect.~\ref{chanresults}.}
\label{FigIntimg}
\end{figure*}

\section{\integral results}
\label{intresults}

A source detected at 15.6$\sigma$ significance in the 20--100\,keV\footnote{We use the 20--100\,keV band for source detection and to quote fluxes, while our spectra are plotted in the 20--200\,keV range to show their high energy spectral dependence.} energy band in the 3$^{rd}$ IBIS (Imager on Board \integral Satellite) catalogue \citep{BirdMaliziaBazzano2007} has been associated with \hbox{PSR~J1846$-$0258} and/or its PWN.  The data from this survey cover the epoch from the launch of \integral (2002 October) up until 2006 May with a total exposure of 1.8\,Ms in this region of the sky.  Individual pointings processed using the \integral Offline Science Analysis v.5.1 (OSA, \citealt{GoldwurmDavidFoschini2003}) were mosaicked to create an all sky image according to the processes described in \citet{BirdMaliziaBazzano2007}.  This allowed the detection of sources as faint as \hbox{PSR~J1846$-$0258}, which is detected at position $\alpha_{2000}=18^{\rm h}46^{\rm m}23^{\rm s}$, $\delta_{2000}=-2^\circ58'59''$ with a 90\% positional uncertainty of $1.8'$ \citep{GrosGoldwurmCdolle-Bell2003}.  This is $41\arcsec$ from the pulsar position as determined from \chandra observations \citep{HelfandCollinsGotthelf2003}.  Figure~\ref{FigIntimg} shows the detection of \hbox{PSR~J1846$-$0258} with IBIS/ISGRI (\integral Soft Gamma-ray Imager, \citealt{UbertiniLebrunDiCocco2003}) in a significance map generated in the 20--100\,keV energy range. 

A spectrum was generated by calculating the weighted mean flux over the entire IBIS/ISGRI exposure time at the source position in a number of narrow energy bands.    Spectral analysis was performed with XSPEC v.11.3.2 \citep{Arnaud1996} using systematic errors of 2\%. The best fit was achieved with a power law model of slope $\Gamma=2.0\pm0.2$ 
with $\chi^2{\rm(dof)}=9.4(13)$.  Absorption does not affect the spectrum significantly at this energy range, and fixing it at $N_{\rm H}=3.96\times10^{22}$\,cm$^{-2}$ \citep{HelfandCollinsGotthelf2003} did not alter the fit.  The flux in the 20--100\,keV energy range is $2.9^{+0.2}_{-0.1}\times10^{-11}$\,erg\,cm$^{-2}$\,s$^{-1}$, corresponding to a luminosity of $1.3\times10^{36}$\,erg\,s$^{-1}$ at a distance of 19\,kpc. 

The source is also detected at the 3.3$\sigma$ level in 87\,ks of exposure time in JEM-X (Joint European X-ray Monitor, \citealt{LundBudtzWestergaard2003}).  All science windows with a pointing position within $2.4^\circ$ of the position of \hbox{PSR~J1846$-$0258} were used to generate a mosaic in the 3--26\,keV band.  The flux can be estimated as $\sim2.6\times10^{-11}$\,erg\,cm$^{-2}$\,s$^{-1}$, corresponding to a luminosity of $\sim1.1\times10^{36}$\,erg\,s$^{-1}$ (at 19\,kpc) but insufficient statistics to allow the generation of a spectrum are available.

The timing properties of the $\gamma$-ray source were not investigated due to the short period and low detection significance in both instruments.   

\section{\chandra Observations}
\label{chanresults}

In 2000 October \citet{HelfandCollinsGotthelf2003} performed imaging spectroscopy of the region around \hbox{PSR~J1846$-$0258} with the ACIS camera aboard {\it Chandra}.  The 1.0--7.0\,keV image from their work shows the SNR of Kes 75 with a bright centrally located point source identified with the pulsar and diffuse emission from the PWN (the pulsar and PWN are shown on the inset of Fig.~\ref{FigIntimg}).  Clear structure is visible in the PWN with the emission arising predominantly around an axis through the pulsar, at an inclination 30$^\circ$ east of north.  There are two bright spots along this axis:  one on either side of the pulsar.  \citet{HelfandCollinsGotthelf2003} were able to extract spectra for all three components.  They used a thermal bremsstrahlung model with $kT=2.9\pm0.1$\,keV to fit the SNR, an absorbed power law with slope $\Gamma=1.39\pm0.04$ to fit the pulsar and an absorbed power law model with a slope of $\Gamma=1.92\pm0.04$ for the PWN.  The authors also investigated the spectral shape over different morphological regions of the PWN and found the power law slopes of all regions to agree with the value quoted above to within 2$\sigma$.  

Due to the fact that IBIS/ISGRI does not resolve the supernova, pulsar and PWN as separate components, we employed \chandra spectral information to assist in clarifying the source of emission of the high energy photons.  We reprocessed the archival \chandra data \citep{HelfandCollinsGotthelf2003} with CIAO~3.4, correcting the level 1 event data for charge transfer inefficiency and filtering for good event grades and good time intervals by excluding bright background flares.  The resulting exposure was 35\,ks.  We extracted spectra from 4 spatial regions, using the CIAO scripts {\it specextract} for extended sources and {\it psextract} for point sources. 

\begin{figure}
 \resizebox{\hsize}{!}{\includegraphics{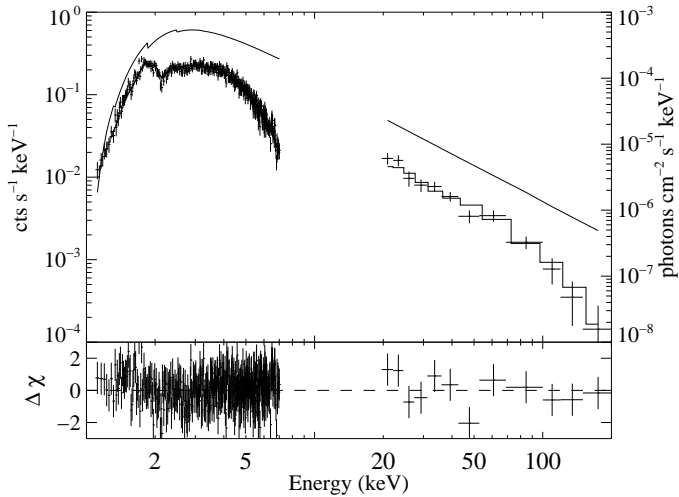}}
 \caption{Combined \chandra and IBIS/ISGRI spectrum of the PWN and \hbox{PSR~J1846$-$0258} in the energy range 1--200\,keV.  In the top panel the data are shown as crosses, the model fit as the histogram and the solid line is the photon spectrum as plotted on the right y-axis.  The lower panel shows the residuals.}
  \label{FigSpec}
\end{figure}

Fitting these various \chandra components alongside the IBIS/ISGRI data for \hbox{PSR~J1846$-$0258}, we can evaluate the contribution of these components to the flux and the spectral slope in the soft \gm ray energy range.  For \chandra data we fit in the 1--7\,keV range where the statistics are most reliable, and extrapolate the fluxes to the 0.5--10\,keV range so that they can be easily compared with other sources in the literature.  We can summarise as follows:
\begin{enumerate}
 \item [{\small (i)}] The entire SNR (a circle of radius $100\arcsec$ centred on, but excluding, the pulsar and the PWN):  The flux from the SNR dominates the emission of the system between 0.5--3\,keV, but being of thermal origin with a temperature $\sim 3$\,keV, quickly becomes too weak relative to the other components of the system to contribute significantly to the high energy emission detected by IBIS/ISGRI.
 \item [{\small (ii)}] The PWN:  Fitting the \chandra data along with the IBIS/ISGRI data results in a power law slope of $\Gamma=1.84\pm0.02$ and $N_{\rm H}=(3.9\pm0.1)\times10^{22}$\,cm$^{-2}$ with $\chi^2_\nu{\rm(dof)}=0.9(369)$.
 \item[{\small (iii)}] The pulsar:  Fixing the absorption at the value derived from our fit to the \chandra data from the PWN ($N_{\rm H}=(3.9\pm0.1)\times10^{22}$\,cm$^{-2}$), we find that the combined \chandra and IBIS/ISGRI spectrum can be fit with a power law of slope $\Gamma=1.63^{+0.08}_{-0.17}$ where $\chi^2_\nu{\rm(dof)}=1.2(161)$.  Although this fit is adequate, as constrained by the statistics in the \chandra region, the residuals in the IBIS/ISGRI energy range indicate that a steeper slope would better describe the high energy data.
 \item[{\small (iv)}] Pulsar + PWN:  As above, the absorption was fixed to $N_{\rm H}=(3.9\pm0.1)\times10^{22}$\,cm$^{-2}$, and the combined \chandra and IBIS/ISGRI spectra could be fit with a power law of slope $\Gamma=1.82\pm0.03$ where $\chi^2_\nu{\rm(dof)}=0.97(380)$.  This spectrum is plotted in Fig.~\ref{FigSpec} and the spectral parameters given in Table~\ref{TabSpecPar}.
\end{enumerate}

\begin{figure}
 \resizebox{\hsize}{!}{\includegraphics{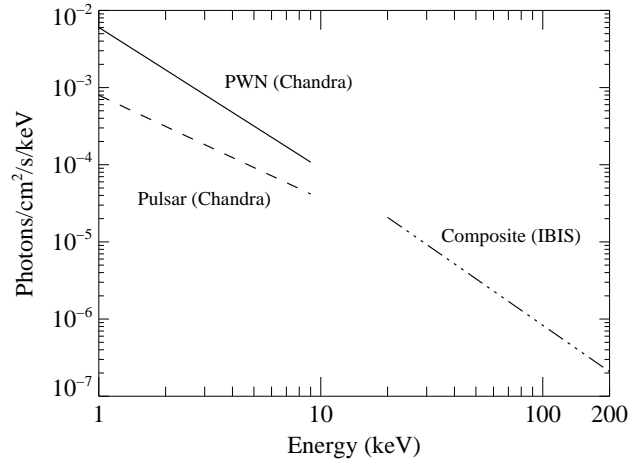}}
 \caption{Models of the pulsar (dashed line) and PWN (solid line) as determined from the \chandra data alongside the model of IBIS/ISGRI composite spectrum (dash-dotted line).  The higher flux from the PWN in the 20--60\,keV band causes it to dominate the statistics and hence determine the measured spectral slope over the whole 20--200\,keV range.}
 \label{FigCompare}
\end{figure}

The model which excludes the PWN emission {\small(iii)} fails to provide a good fit to the IBIS/ISGRI data.  Conversely, as can be seen by comparing points {\small(ii)} and {\small (iv)} above,  including the pulsar contribution has very little effect on the spectral slope.  As we have no data which can distinguish between the pulsar and PWN components above 10\,keV we cannot rule out spectral breaks in either of the spectra above this energy.  Although it is possible that both the pulsar and PWN spectra could suffer breaks such that their combined spectrum gives rise to the spectrum we detect in the IBIS/ISGRI energy range, the simplest scenario is that the PWN, which dominates the flux at 10\,keV, still dominates at 20\,keV.  Despite the extrapolated pulsar spectrum becoming dominant beyond 50\,keV, the measured IBIS/ISGRI spectral slope is then that of the PWN, as the spectral slope is determined in the 20--60\,keV region where the statistics are best.  This is supported by the fact that the slope of the PWN, as measured from \textit{Chandra} data, is consistent with that of the composite IBIS/ISGRI spectrum.  Figure~\ref{FigCompare} illustrates this with models of the \chandra and IBIS/ISGRI spectra.

\begin{table}
\caption{Spectral parameters for the combined \chandra and IBIS/ISGRI fit to the region around \hbox{PSR ~J1846$-$0258} (excluding SNR from the \chandra data). $C$ is a normalisation parameter introduced to account for uncertainties in the cross-calibration of \chandra and IBIS/ISGRI.  The unabsorbed luminosity values are given assuming a distance of 19\,kpc.}
\label{TabSpecPar}
\begin{tabular}{l l}
\hline\hline
\smallskip
Parameter & Value\\
\hline
$N_{\rm H}$(fixed) & $3.9\times10^{22}$\,cm$^{-2}$\\
\smallskip
$\Gamma$ & $1.82\pm0.03$\\
\smallskip
$C$ & $0.8\pm0.1$ \\
\smallskip
$\chi^2$(d.o.f.) & 369(380)\\
\smallskip
$L_{X}$(0.5--10\,keV) & $1.7\times10^{36}$\,erg\,s$^{-1}$\\
\smallskip
$L_{\gamma}$(20--100\,keV) & $1.3\times10^{36}$\,erg\,s$^{-1}$\\
\hline
\end{tabular}
\end{table}

\section{Discussion}
\label{discuss}

PSR~J1846$-$0258 is one of a recently emerged and growing class of rotation-powered pulsars with inferred surface dipole magnetic fields approaching those of the magnetars.  A number of these pulsars have been discovered as a result of the Parkes multi-beam pulsar survey of the Galactic plane \citep{ManchesterLyneCamilo2001}, for example \hbox{PSR~J1119$-$6127} and \hbox{PSR~J1814$-$1744} \citep{CamiloKaspiLyne2000}, which have inferred magnetic fields of $4.1\times10^{13}$\,G and $5.5\times10^{13}$\,G respectively.  These are all young pulsars with characteristic ages of order a few thousand years and approximately half of them show X-ray emission (\hbox{PSR~J1119$-$6127}, \citealt{GonzalezSafi-Harb2003}; \hbox{PSR~J1718$-$3718}, \citealt{KaspiMcLaughlin2005}) with spectra harder than those of magnetars.  

In fact, \hbox{PSR~J1119$-$6127} shows a remarkable resemblance to \hbox{PSR~J1846$-$0258} as far as spin characteristics and the properties inferred from them are concerned.  \hbox{PSR~J1119$-$6127} has $P=408$\,ms and $\dot{P}=4.1\times10^{-12}$\,s\,s$^{-1}$, giving a spin-down luminosity, characteristic age and inferred surface dipole magnetic field very close to those of \hbox{PSR~J1846$-$0258} \citep{CamiloKaspiLyne2000}.  This, however, is where the similarity ends.  Across the electromagnetic spectrum, these pulsars have very different attributes.  In the soft $\gamma$-ray band covered by IBIS/ISGRI, \hbox{PSR~J1846$-$0258} is clearly detected in the 20--100\,keV band (see Table~\ref{TabSpecPar}), while with similar exposure we can place a $2\sigma$ upper limit of $3.3\times10^{33}$\,erg\,s$^{-1}$ on any emission from the location of \hbox{PSR~J1119$-$6127} (using a distance of  8.4\,kpc, \citealt{CaswellMcClure-GriffithsCheung2004}) in the same energy band.  Also in the X-ray band (0.5--10\,keV), these two pulsars show contrasting spectral behaviour.  While \hbox{PSR~J1846$-$0258} shows absorbed power law spectra from both the pulsar and PWN \citep{HelfandCollinsGotthelf2003}, \hbox{PSR~J1119$-$6127} shows a strong thermal spectrum superimposed on a hard power law \citep{GonzalezKaspiCamilo2005}.  Although \citet{GonzalezKaspiCamilo2005} attribute the hard power law component to the PWN, the thermal X-rays from the pulsar are difficult to interpret through conventional emission models, e.g. models for initial cooling of the entire neutron star result in a radius smaller than allowed by neutron star equations of state, while hot spot models result in unusually high temperature estimates.  It may be the case that, while these two pulsars show similar current spin characteristics, their evolutionary paths and physical properties may be very different, as reflected in the significantly different emission properties.

The most unusual feature of \hbox{PSR~J1846$-$0258} is its efficiency in converting spin-down power, $\dot E$, into X- and $\gamma$-ray luminosity $L_{\rm X}$ and $L_{\rm \gamma}$.  Using data from 41 pulsars, \citet{PossentiCeruttiColpi2002} determined a general relation between the 2--10\,keV X-ray luminosity of the combined pulsar and PWN, and the spin-down power.  According to this relation, and the value of $\dot E$ determined from the spin characteristics, $L_{\rm 2-10\,keV}/\dot{E}$ should be $\sim 0.2$\% for \hbox{PSR~J1846$-$0258}.  The measured efficiency, at $L_{\rm 2-10\,keV}/\dot{E}\sim 12$\%, when combined with the 20--100\,keV measured in this work indicates $L_{\rm 2-10,20-100\,keV}/\dot{E}\sim27$\%.  It is expected that the total efficiency in converting spin-down power into luminosity is actually much larger than this, as the luminosity contribution between 10--20\,keV is not included and the soft $\gamma$-ray flux extends beyond 100\,keV (see Fig.~\ref{FigSpec}).  The observation that the 2--10\,keV spin-down conversion efficiency is so much larger than average, and the fact that this efficiency  becomes even larger upon inclusion of the 20--100\,keV luminosity, may be a reason to call the distance estimate into question.

To this effect, we can compare the efficiency of \hbox{PSR~J1846$-$0258} with two similar young pulsars in the Large Magellanic Cloud, for which the distance is well known at $\sim50$\,kpc.  Including luminosity contributions from both the pulsar and PWN \hbox{PSR~J0537$-$6910} has $L_{\rm 0.5-10\,keV}/\dot{E}=0.9$\% \citep{ChenWangGotthelf2006} and \hbox{PSR~B0540$-$69} has $L_{\rm 0.5-10\,keV}/\dot{E}=9.1$\% \citep{KaaretMarshallAldcroft2001}. Even though the estimate for \hbox{PSR~B0540$-$69} is much larger than the X-ray efficiency of most pulsars, it is still only around half that of \hbox{PSR~J1846$-$0258} ($L_{\rm 0.5-10\,keV}/\dot{E}=20$\%).  \citet{BeckerHelfand1984} determined a distance of 19\,kpc through neutral hydrogen measurements.  This distance, which agrees with the estimate of \citet{Milne1979} based on the surface brightness -- diameter relation for supernovae and the neutral hydrogen measurements of \citet{CaswellMurrayRoger1975}, does imply a very high mean expansion velocity for the supernova remnant \citep{HelfandCollinsGotthelf2003}.  On the other hand, if the distance were overestimated, this would lead to inconsistency between the neutral hydrogen column as measured by \citet{BeckerHelfand1984} and that deduced from fits to X-ray data of the PWN \citep{HelfandCollinsGotthelf2003} -- such inconsistency is not observed.  \citet{MortonSlaneBorkowski2007} also discuss the possibility of an incorrect distance estimate, but dismiss this option as it would double the already high inferred density of postshock gas in the SNR.  We can estimate a lower limit on the distance of the system by employing the half width at half maximum of the Si line as measured by \citet{HelfandCollinsGotthelf2003}.  Assuming that the SNR has been expanding at this measured rate (1850\,km\,s$^{-1}$) for its entire lifetime (an upper limit of 884 years, \citealt{LivingstonKaspiGotthelf2006}), we can estimate the distance from the angular size of the SNR on the sky to be $\sim3$\,kpc.  The real distance is certainly larger than this lower limit due to the fact that the SNR expansion velocity at earlier times was most likely larger than our current estimate.  Using this lower limit on the distance would provide $L_{\rm 0.5-10\,keV}/\dot{E}\sim 0.5$\% -- more in line with the efficiencies of other rotation-powered pulsar systems.   This makes the prospect of a distance smaller than 19\,kpc to \hbox{PSR~J1846$-$0258} appealing, but difficult to substantiate in terms of the existing estimates \citep{CaswellMurrayRoger1975,Milne1979,BeckerHelfand1984}.

Of the rotation-powered pulsars that exhibit X-ray emission, only very few have been detected above 10\,keV.  Although a consistent treatment exploring the general properties of PWN in the 20--100\,keV range will be presented as a future publication, we can see from the existing literature that the \integral spectrum of \hbox{PSR~J1617$-$5055}, a young rotation powered pulsar, shows a power law slope of $\sim 2$ \citep{LandiDeRosaDean2007}, while a combined IBIS/ISGRI and BeppoSAX spectrum of the PWN in \hbox{PSR~B1509$-$58} \citep{ForotHermsenRenaud2006} can be fit with a power law of photon index $\sim2.1$ -- both close to the spectral shape of \hbox{PSR~J1846$-$0258}.  In addition, the EGRET instrument aboard the Compton Gamma-Ray Observatory detected emission from $\sim7$ rotation-powered pulsars and/or their PWN in the 30\,MeV--20\,GeV energy range \citep{Roberts2005}. These pulsars have photon indices harder than 2.2, also consistent with the soft \gm ray spectral slope of \hbox{PSR~J1846$-$0258} in this work and with those \integral observations quoted above.  However, smooth spectral coverage above 100\,keV and into the GeV energy range is required to ascertain the presence or absence of spectral turnovers predicted by models of PWN emission mechanisms \citep{ZhangHarding2000,ChengHoRuderman1986a}.

\section{Summary}
 With an angular resolution of $12\arcmin$, the IBIS/ISGRI image does not resolve the components of the supernova, pulsar and PWN in the \hbox{PSR~J1846$-$0258} system.  We have thus used the spectral properties of the X-rays and soft \gm rays to find that, assuming there are no spectral breaks above 10\,keV in the spectra of either the pulsar or PWN, the spectral shape of the measured soft \gm ray spectrum is influenced predominantly by the PWN.  The SNR remnant, which has a bremsstrahlung spectrum, may contribute somewhat in the low end of the 20--200\,keV range, but is not the dominant source of the high energy emission.

We have discussed the properties of \hbox{PSR~J1846$-$0258} in the context of other high B-field rotation-powered pulsars, and in particular, \hbox{PSR~J1119$-$6127}, which has very similar spin characteristics.

On the basis of the high observed X-ray luminosity to spin-down power ratio, we explore the possibility of an overestimate in the distance to \hbox{PSR~J1846$-$0258}.  Although a smaller distance could bring this ratio in line with those observed in other rotation-powered pulsars, it results in incongruities in both measurements of the line of sight hydrogen density and in estimates of the postshock density of the SNR.

The increasing availability of soft $\gamma$-ray data from \integral allows us to characterise the emission from rotation-powered pulsars in a previously poorly explored energy range.  Widening the spectral coverage for these systems will help to provide a means to understanding the processes and emission mechanisms involved.

\begin{acknowledgements}
Based on observations with {\it INTEGRAL}, an ESA project with instruments and science data centre funded by ESA member states (especially the PI countries:  Denmark, France, Germany, Italy, Switzerland, Spain), Czech Republic and Poland, and with the participation of Russia and the USA.  University of Southampton authors acknowledge funding from the PPARC grant PP/C000714/1.  INAF/IASF--Roma and INAF/IASF--Bologna authors acknowledge funding from ASI-INAF I/023/05/0 and ASI I/008/07/0.  This research has made use of data obtained through the High Energy Astrophysics Science Archive Research Center Online Service, provided by the NASA/Goddard Space Flight Center.
\end{acknowledgements}

\bibliographystyle{aa}

\label{lastpage}

\end{document}